\documentclass[sigconf,nonacm]{acmart}

\usepackage{graphicx}
\usepackage{booktabs}
\usepackage{multirow}
\usepackage{amsmath,amsthm}
\usepackage{xcolor}
\usepackage{subcaption}
\usepackage{array}
\usepackage{algorithm}
\usepackage{algorithmic}

\newtheorem{proposition}{Proposition}

\setcopyright{none} 
\renewcommand\footnotetextcopyrightpermission[1]{}

\title{Session-Bound Behavioral Watermarking for Autonomous LLM Agents}

\author{Md Facklasur Rahaman}
\affiliation{%
  \institution{Deakin University}
  \city{Melbourne}
  \state{VIC}
  \country{Australia}}
\email{s226124006@deakin.edu.au}

\author{Yong Xiang}
\affiliation{%
  \institution{Deakin University}
  \city{Melbourne}
  \state{VIC}
  \country{Australia}}
\email{yong.xiang@deakin.edu.au}

\author{Md Palash Uddin}
\affiliation{%
  \institution{Deakin University}
  \city{Melbourne}
  \state{VIC}
  \country{Australia}}
\email{m.uddin@deakin.edu.au}

\author{Shang Gao}
\affiliation{%
  \institution{Deakin University}
  \city{Melbourne}
  \state{VIC}
  \country{Australia}}
\email{shang.gao@deakin.edu.au}

\author{Guang Hua}
\affiliation{%
  \institution{Singapore Institute of Technology}
  \city{Punggol Coast}
  \country{Singapore}}
\email{guang.hua@singaporetech.edu.sg}

\author{Leo Yu Zhang}
\affiliation{%
  \institution{Griffith University}
  \city{Gold Coast}
  \country{Australia}}
\email{leo.zhang@griffith.edu.au}

\begin{document}


\begin{abstract}

Large Language Model (LLM) agents are increasingly acting through Web interfaces and external tools. They produce trajectories. Those trajectories may pass through several intermediaries. They can store, modify, or dispute those trajectories. In Web-based agent deployments, trajectories may pass through APIs, marketplaces, and tool servers. As such, a copied or disputed log may not by itself reveal which agent owner produced it. Behavioral watermarking can embed ownership evidence in agent action choices. However, existing methods mainly address detection, removal, rewriting, or imitation. They do not directly test whether repeated observation of watermarked trajectories enables cross-session forgery by exploiting learned action preferences. We formulate this learned-forgery threat and propose AgentAnchor, a session-keyed behavioral watermark for owner attribution from trajectory logs. We derive each step key from a secret key, a fresh public session identifier, and semantic states of recent observations, while a guarded digest of the action-policy state diversifies linear coding rows. It embeds parity evidence through a KL-bounded exponential tilt, and a trusted verifier checks a disputed log without rerunning the agent, using wrong-key calibration and abstention.

\end{abstract}






\maketitle
\pagestyle{plain}

\sloppy


\section{Introduction}
\label{sec:introduction}

\begin{figure*}[t]
  \centering
  \begin{subfigure}[b]{0.41\textwidth}
    \includegraphics[width=\linewidth]{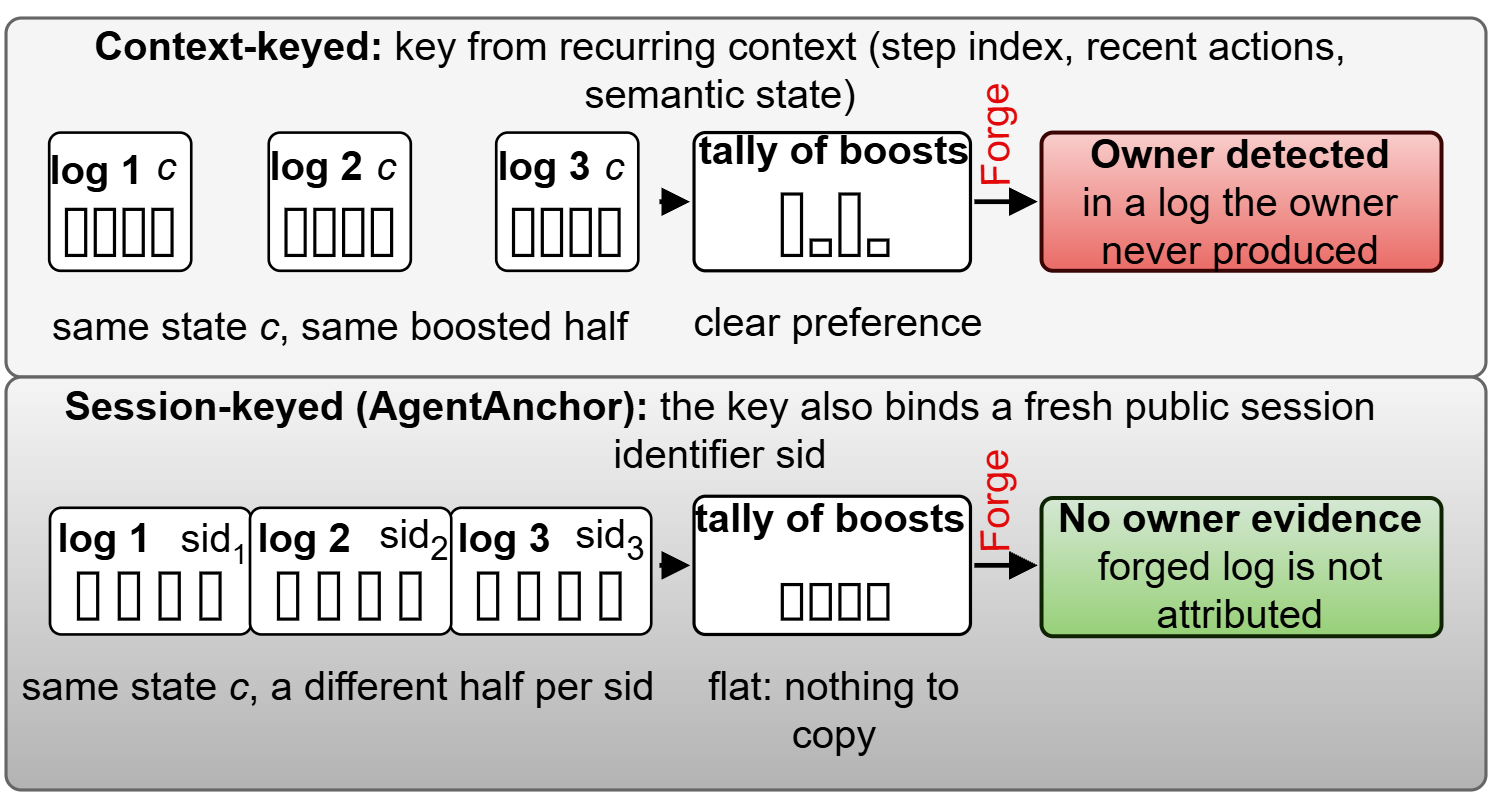}
    \caption{Why a recurring key leaks}
    \label{fig:prob-concept}
  \end{subfigure}\hfill
  \begin{subfigure}[b]{0.28\textwidth}
    \includegraphics[width=\linewidth]{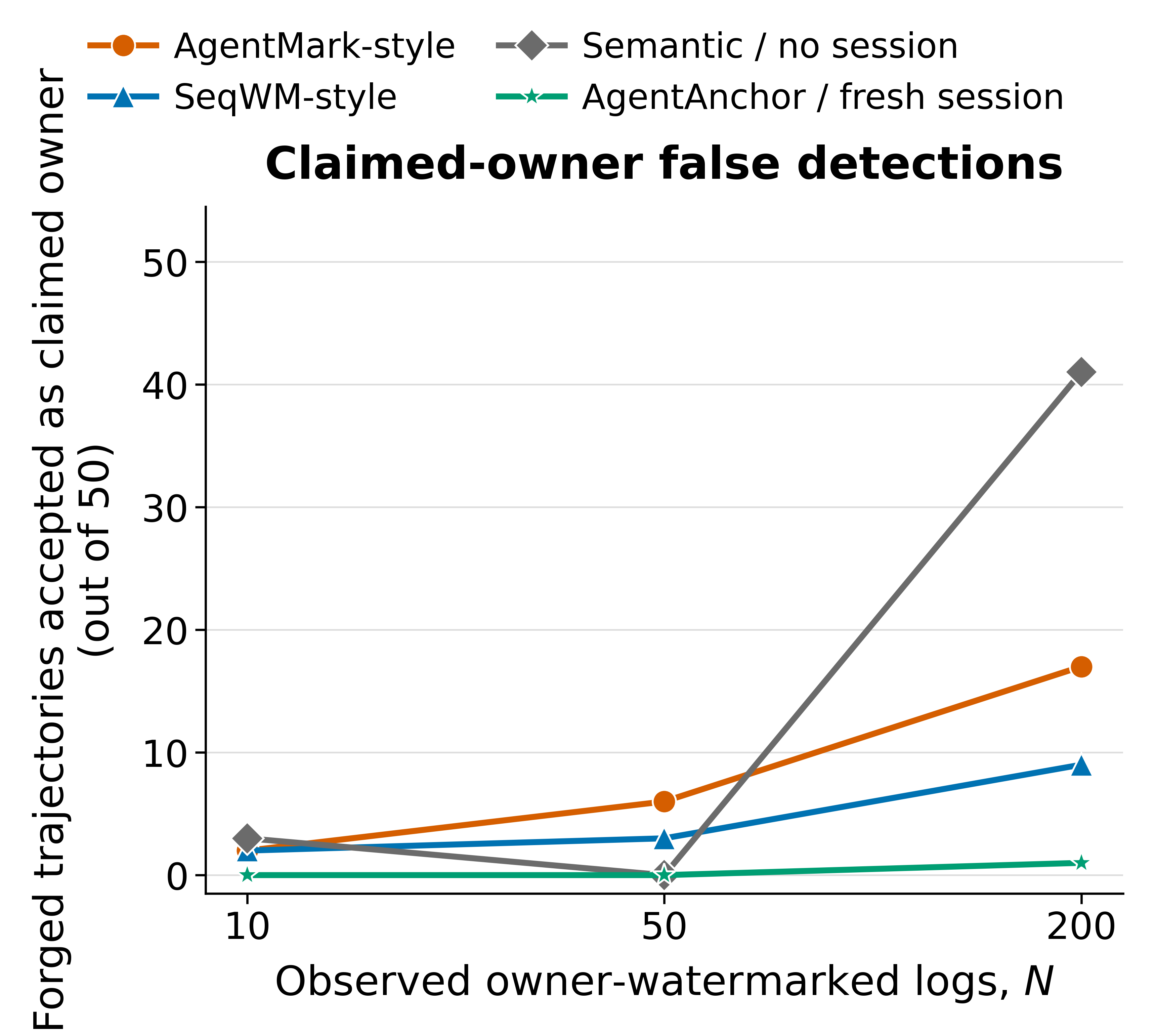}
    \caption{False claimed-owner detection}
    \label{fig:prob-fd}
  \end{subfigure}\hfill
  \begin{subfigure}[b]{0.28\textwidth}
    \includegraphics[width=\linewidth]{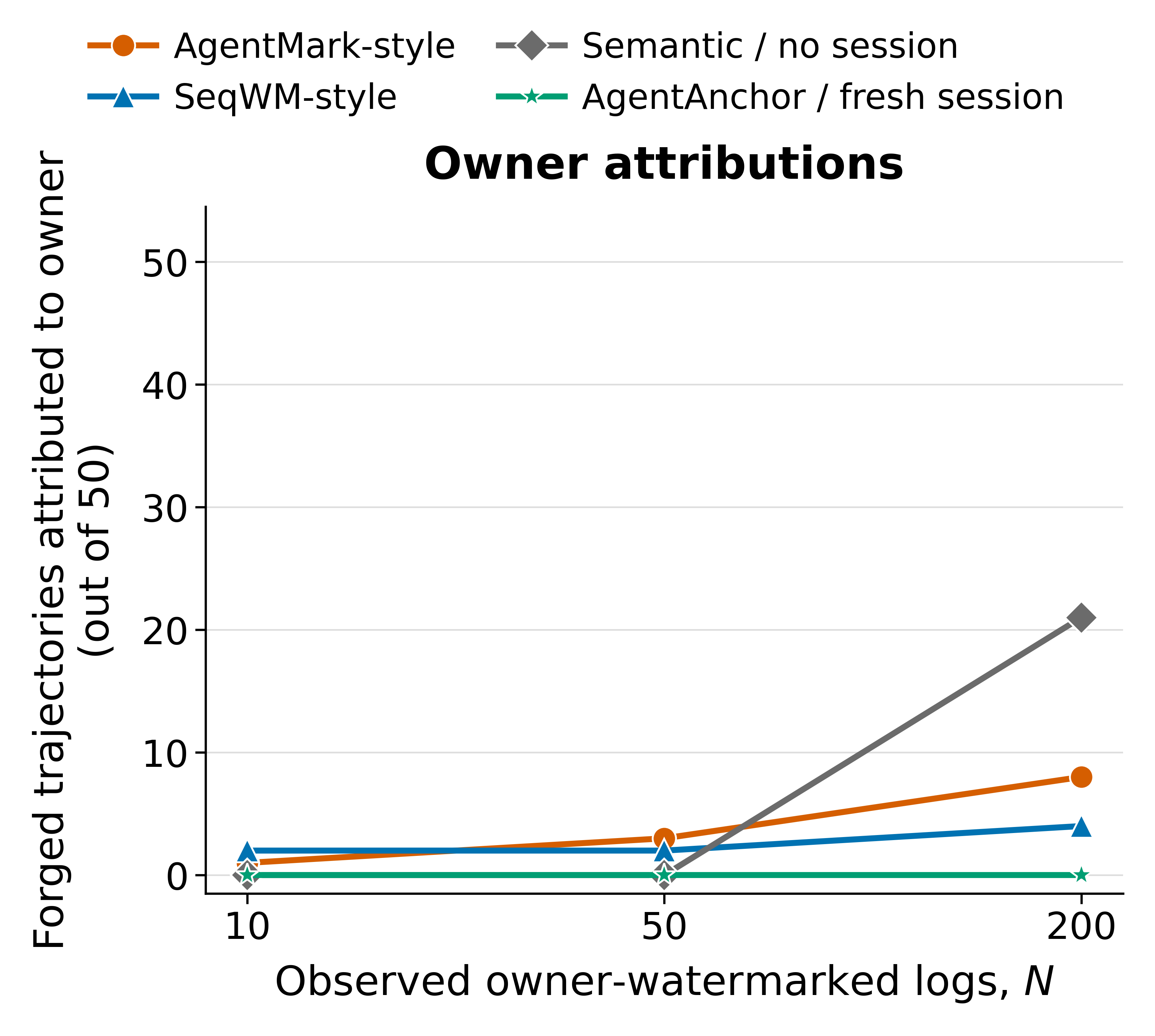}
    \caption{False owner attribution}
    \label{fig:prob-fa}
  \end{subfigure}
    \caption{Learned cross-session forgery. (a) If the watermark key depends on context that recurs across sessions, an observer can learn its keyed preferences from logs and forge trajectories attributed to the owner. Our AgentAnchor binds the key to a fresh session identifier. (b, c) On ALFWorld, forgeries against AgentMark-style~\cite{huang2026agentmark} and SeqWM-style~\cite{an2026sequential} and session-free semantic keys pass more often as more logs are observed, while our AgentAnchor keeps them near zero.}
  \label{fig:problem}
\end{figure*}

Large Language Model (LLM)-based agents are increasingly carrying out users' tasks on the Web. Several benchmarks use real-world data or realistic Web environments to evaluate such agents \cite{yao2022webshop}\cite{deng2023mind2web}\cite{zhou2024webarena}. Deployed agents can also access several external tools through interfaces such as the Model Context Protocol (MCP). These executions produce agent trajectories, which can record the observations received by agents and the actions that the agents take. However, these trajectories may not remain with the agent owner all the time. The trajectory steps can pass through several Application Programming Interface (API) gateways and MCP servers. In those places, the trajectories may be stored, copied, or even modified. These various parties or places cannot always be trusted. In a study of 45 live MCP servers, poisoned tool descriptions caused agents to misuse genuine tools in up to 72.8\% of test cases \cite{wang2026mcptox}.

Consider a scenario where a vendor of a shopping agent provides an API for their marketplace and then disputes a purchase made through that agent. The concerned party, gateway or tool server, can have a copy of the trajectory. However, the log alone does not identify who created the trajectory. A behavioral watermark can address this problem by placing a keyed signal into the action choices performed by the agent. A verifier holding the owner's secret key can later check a copied log for this signal, as long as the copy keeps the actions and logged fields this check needs. There is a downside to using this method as well. The watermarked actions are visible to every party that sees the logs. If an attacker collects enough trajectories from the owner, the attacker may be able to determine what type of actions the watermark prefers in each specific scenario. The attacker may also be able to replicate this type of action in a new trajectory that has never been created by the owner.

Such a forgery may incorrectly attribute a damaging trajectory to the owner. On the Web, this could get an honest vendor blocked or blamed for fraud. This creates a security risk in addition to the ability to remove a watermark (which only renders an original trajectory unattributable). For text watermarks, an attacker who queried a watermarked model approximately learned its rules and spoofed watermarked text for under \$50, with an average attack success rate above 80\% \cite{jovanovic2024watermark}. Researchers have also investigated ways to protect ownership and verify it for other types of online resources such as black-box LLM service providers through fingerprinting \cite{shao2026reading}, certified verification of training datasets \cite{qiao2026dssmoothing}, and graph-based data via watermarking \cite{li2026drgw}. Our focus is on the attribution of agent trajectories under two conditions. Firstly, the verifier relies on the structured logged trajectory and the owner's secret key, without rerunning the agent, and secondly, the parties who observe the logs should not be able to learn the watermark from them.


\subsection{Related Work and Motivation}

Behavioral watermarking embeds ownership evidence directly in an agent's decisions, which supports process-level provenance. AgentMark embeds identifiers through an agent’s behavioral or tool-use choices using distribution-preserving sampling \cite{huang2026agentmark}. ActHook protects trajectory datasets by inserting trigger-activated hook actions that a trained model later reproduces \cite{meng2026watermarking}, while SeqWM uses sequential transition patterns to preserve watermark detectability under trajectory corruption \cite{an2026sequential}. TRACE has two channels, which are a content-keyed action-selection channel for deletion robustness and a structure-based tally channel for rewriting robustness. They are complementary to each other. \cite{gao2026trace}. TrajMark targets coding-agent trajectories by encoding an owner identifier through keyed READ actions and adding seals for local tamper localization \cite{zeng2026trajmark}. These works show that agent behavior and trajectory structure can carry useful provenance signals. Other work protects related but different assets. AGENTWM biases functionally equivalent actions so imitation models can inherit detectable ownership evidence \cite{wang2026protecting}. CoTGuard uses trigger queries to reveal copyright leakage in multi-agent reasoning \cite{wen2025cotguard}, while MemMark embeds attribution signals in long-term memory writes and verifies them with cryptographic commitments and signed session anchors \cite{zhang2026memmark}.

Our concern is with a different type of failure mode. The evaluations of these methods mainly measure detection by the verifier and whether an attacker can delete or rewrite trajectory records, imitate the agent, or tamper with the log. They do not directly test whether a party that observes many prior trajectories can learn enough about the watermark to forge a new trajectory that is falsely attributed to the owner. Several prior trajectory watermarks derive their randomness from contexts that may recur across runs, such as the step index \cite{huang2026agentmark}, previous actions \cite{an2026sequential}\cite{gao2026trace}, or file names \cite{zeng2026trajmark}, so the same keyed preference recurs whenever that context recurs. Watermark behavior is also learnable from past outputs. For example, watermark distillation can train a model to reproduce text watermarks \cite{gu2024learnability}, and watermark stealing can approximate secret watermark rules well enough to spoof new text \cite{jovanovic2024watermark}. Figure~\ref{fig:problem} shows that the same threat is measurable for agent trajectories. In a post-hoc replay of our logged ALFWorld trajectories, we re-embedded the watermark under different step-key inputs while keeping everything else fixed, let an attacker learn keyed preferences from $N$ observed watermarked logs, and used them to rewrite 50 unwatermarked trajectories (Appendix~\ref{app:keying}). With $N=200$, forgeries passed claimed-owner detection for 17, 9, and 41 of the 50 targets when the key used the step index with recent actions (AgentMark-style), the preceding actions (SeqWM-style), or our semantic state without a session identifier (Figure~\ref{fig:prob-fd}), and 8, 4, and 21 were attributed to the owner (Figure~\ref{fig:prob-fa}). When the key also bound a fresh session identifier, only 1 forgery passed detection and none was attributed. This motivates the primary research question that we will address in this paper: \textit{Can a watermark remain verifiable from structured trajectory logs while limiting what an observer can transfer from previously observed sessions to a new one?}


\subsection{Our Approach and Contributions}

We propose AgentAnchor, a session-keyed behavioral watermark for owner attribution from logged agent trajectories. Our primary idea is to bind every step key to a fresh session. So, keyed preferences learned from earlier sessions do not transfer to a new one, while semantic windows keep the key stable under renumbering and tolerant of moderate rewording. At each step, we score the admissible actions with the agent's own model to obtain an action distribution. We derive the step key using an HMAC over the owner's secret, a fresh public session identifier, and the semantic clusters of recent observations. The step key and a perturbation-tolerant digest of the most probable actions then derive a coding row and one parity bit of the owner's payload. We embed this bit through a KL-bounded exponential tilt of the action distribution toward one half of a keyed action split. During verification, we recompute the keys from the structured log and the owner's secret, decode the payload from the logged actions and probabilities, and calibrate the evidence against many wrong keys. We accept attribution only when the evidence is significant and supports a unique owner payload. Otherwise, we abstain. AgentAnchor builds on random linear payload coding from AgentMark \cite{huang2026agentmark} and the exponential tilt and wrong-key calibration used in SeqWM \cite{an2026sequential}. MemMark also puts dialogue identifiers in its keys, but it watermarks memory writes and does not test whether behavior learned from observed sessions can be forged into a new one \cite{zhang2026memmark}. Unlike distribution-preserving approaches \cite{huang2026agentmark}, \cite{gao2026trace}, \cite{zhang2026memmark}, our tilt changes the action distribution, and our verifier relies on the logged action probabilities. We also assume that the deployment keeps the session identifier fresh and unaltered, for example by issuing it from the verifier. Our contributions are as follows.


\begin{itemize}

\item We develop a session-keyed behavioral watermark for LLM agent trajectories that separates semantic synchronization from multi-bit coding diversity. Semantic clusters of recent observations make the key independent of step numbers and tolerant of moderate rewording, while a perturbation-tolerant digest of the most probable actions provides diverse coding rows for owner identification.

\item We study learned forgery of behavioral watermarks from previously observed trajectories. Deliberate session reuse creates transferable watermark behavior, while fresh session binding sharply limits cross-session forgery under the tested attackers. We have evaluated our AgentAnchor framework on ALFWorld (50 tasks) and on the 1,000-product subset of WebShop (WebShop-1K), a simulated e-commerce website.

\end{itemize}


\section{Problem Statement and Design Goals}
\label{sec:problem}


\subsection{System Model}
\label{sec:system-model}

\paragraph{Agent and trajectory.}
We consider an owner who deploys an LLM agent to act on behalf of users. An episode begins with a goal $g$ and lasts for at most $T_{\max}$ steps. At step $t$, the agent receives a textual observation $o_t$, such as the environment's response or the current page text, and a finite admissible action set $A_t$, selects an action $a_t \in A_t$, and receives the next observation $o_{t+1}$. The resulting sequence of observations and actions is the agent's trajectory. We require $A_t$ to be enumerable. This holds directly in text environments such as ALFWorld \cite{shridhar2020alfworld} and for Web interactions whose current page exposes links, buttons, and form actions. For free-form inputs such as WebShop search text \cite{yao2022webshop}, the agent model writes one search query greedily, without the key, and this query becomes the only admissible action. Because the owner controls the agent, it can score every admissible action with the agent's own model and construct an operational action distribution $P_t$ over $A_t$. Without a watermark, the agent samples its action from $P_t$, which is the baseline.


\paragraph{Trajectory log.}
Each decision step produces a public record containing the record index, task identifier, goal, observation $o_t$, admissible set $A_t$, fixed-point representation of $P_t$, selected action $a_t$, a public session identifier $\mathit{sid}$, the public codebook version, and the validity and source fields associated with $P_t$. We call the ordered sequence of these records a \emph{trajectory log}. Each trajectory forms one session, identified by a 128-bit public value $\mathit{sid}$ that stays constant within the trajectory and should be fresh across trajectories. It must be unique but need not be secret or unpredictable. Our runs derive it by hashing the run, task, and counter. The owner logs $P_t$ so that the verifier can score actions without the agent model. The log contains no secret key $\kappa$, no watermark flag, and none of the agent's planning notes.


\paragraph{Parties and secrets.}
The embedding agent and verifier share a secret key $\kappa$. Each registered owner is represented by an $L$-bit identifier $m \in \{0,1\}^L$, which the watermark carries as its payload. Our implementation uses $L=8$, giving $2^L=256$ candidate identifiers. The verifier may be the owner or another party trusted with $\kappa$. Since a holder of $\kappa$ can generate watermark evidence for any payload, we assume that parties entrusted with $\kappa$ are honest.


\paragraph{Verifier.}
The verifier receives $\kappa$ and either one trajectory log or several logs claimed to carry the same identifier. It also has the fixed public AgentAnchor configuration. It does not rerun the agent and does not access the agent model. The verifier may attribute the evidence to a unique decoded identifier $\hat m$, detect evidence for an identifier $m_0$ named before verification (for example, the owner claimed in a dispute) without unique attribution, or abstain.


\subsection{Threat Model}
\label{sec:threat-model}

\paragraph{Adversary.}
We consider a party that can observe trajectory logs produced by the owner's agent, such as an intermediary on the logging path. Following Kerckhoffs's principle \cite{kerckhoffs1883cryptographie}, the adversary knows the embedding and verification algorithms, sentence encoder, semantic codebook, canonicalization rules, and all public parameters. It does not know $\kappa$, and we do not give it oracle access to the verifier. For the learned-forgery experiment, we also give it the agent's unwatermarked logs on the target tasks, with their logged action distributions, which it edits into forgeries.


\paragraph{Adversary goals.}
We consider two goals. The first goal is \emph{watermark removal}, which modifies an authentic watermarked log so that the verifier no longer attributes it, while the resulting log remains usable as a trajectory. Our evaluation includes step renumbering, interface padding, probability re-quantization and perturbation, observation paraphrasing and round-trip translation, adaptive rewriting using the public codebook, step erasure, insertion, truncation, reordering and substitution, action renaming, and removal or replacement of session context.


The second goal is \emph{watermark forgery}, which presents a trajectory that the owner did not produce and attempts to make the verifier attribute it to that owner. The implemented forgers observe $N$ watermarked trajectories, estimate action preferences from public states, and edit the selected actions of an unwatermarked trajectory, which we call the target. In the fresh-session condition, the target carries a session identifier absent from all observed trajectories. We also evaluate a session reuse condition where the attacker has already observed the target's session. This condition violates our freshness assumption below and measures the effect of losing session separation. Because attribution indicates that a log contains owner-generated watermarked steps, not that every action in it is authentic, we treat edited copies of authentic logs as removal attempts and reserve forgery for trajectories without owner-generated steps.


\paragraph{Assumptions and non-goals.}
We make five explicit assumptions and scope restrictions. Firstly, $\kappa$ remains secret. Secondly, session identifiers are fresh and authentic, for example, issued and recorded by the verifier or by the marketplace or gateway that starts each session, so a verifier can check a logged $\mathit{sid}$ against this record. Our AgentAnchor itself does not authenticate logs or prevent replay. A missing, malformed, or inconsistent session identifier causes the verifier to abstain. A different but syntactically valid identifier is processed as a different session and is evaluated as a session-replacement attack. Thirdly, the verifier trusts the logged $P_t$ values as the action distribution used at execution time. This trust matters for attributing genuine logs, but not for the fresh-session bound of Appendix~\ref{app:forgery}, which holds for any logged values. Probability edits are evaluated only empirically as removal attacks. Fourthly, AgentAnchor does not preserve the action distribution exactly. Finally, our empirical forgery results apply only to the implemented attackers.


\subsection{Design Goals}
\label{sec:design-goals}

Our first goal is log-based verification. Verification should use only the structured trajectory log, the secret key $\kappa$, and fixed public AgentAnchor parameters, without rerunning the agent, accessing its model, or reading the agent's planning notes. Our second goal is session-bound keying. The public session identifier should enter the HMAC-based key derivation so that, if HMAC is a pseudorandom function~\cite{bellare2015new}, a fresh session separates its keyed mapping from those of earlier sessions even when the same semantic context reappears. Our third goal is stable synchronization and coding, meaning that the verifier can recompute, from a possibly edited log, the same step keys and coding rows used during embedding. The key and coding-row derivation should therefore avoid direct dependence on record indices, log formatting, raw observation strings, and exact probability values. Our fourth goal is calibrated decisions with abstention. Under the wrong-key exchangeability null, detection and attribution should be calibrated at a declared significance level $\alpha$, and when the available evidence cannot identify one payload, the verifier should abstain rather than guess. Our fifth goal is bounded local distortion. At every embedded step, the modified action distribution should stay within a Kullback--Leibler (KL) divergence budget \cite{kullback1951information} $\eta$ of $P_t$, which bounds how much each step's behavior can change, and should preserve its support.



\section{AgentAnchor Framework}
\label{sec:framework}

\begin{figure*}[t]
    \centering
    \includegraphics[width=\textwidth]{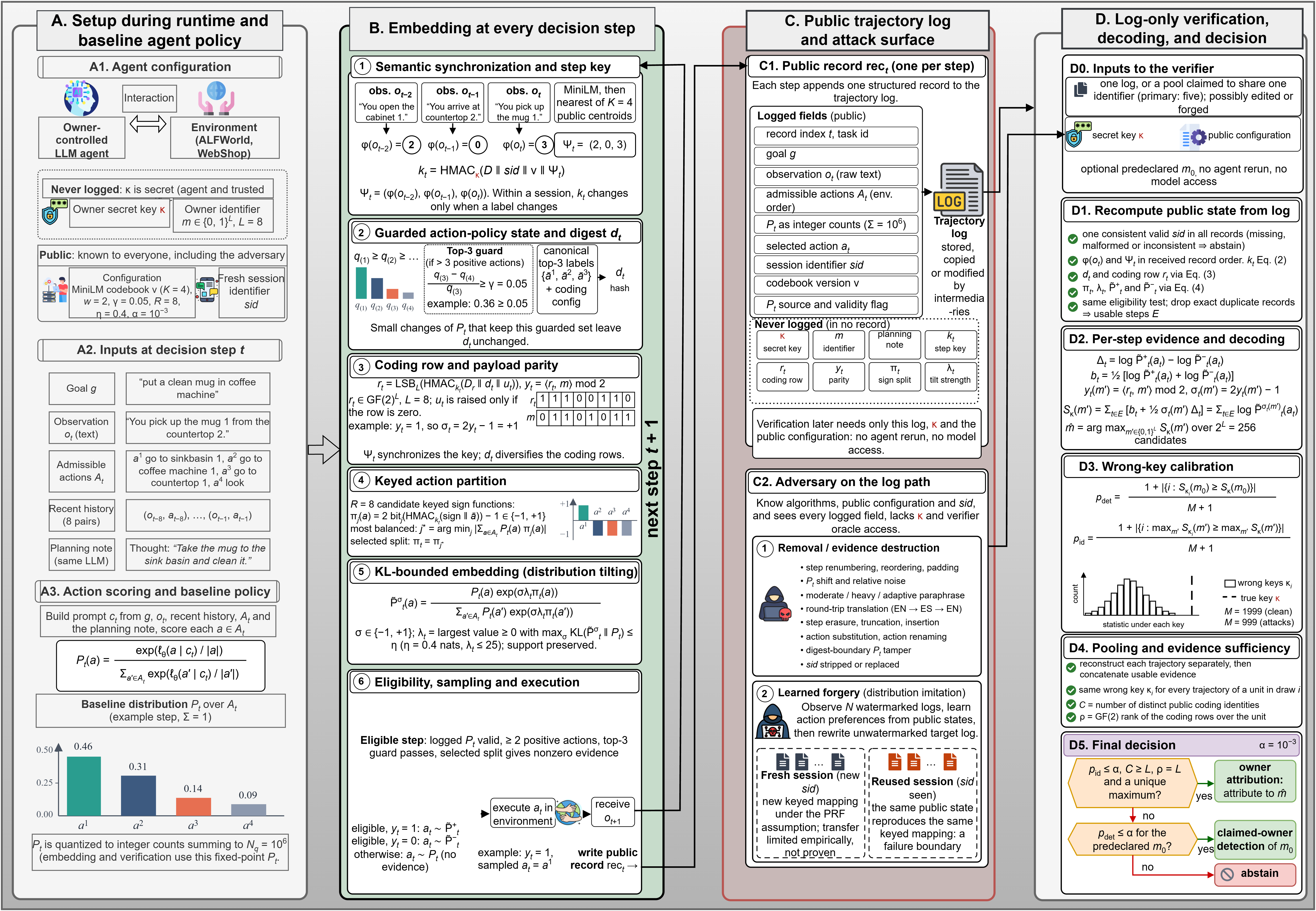}
    \caption{Overview of the AgentAnchor framework}
    \label{fig:framework}
\end{figure*}

Our framework works in two phases. While the agent acts, each eligible step embeds one parity bit of the owner identifier in its action choice, and later a verifier with the secret key collects this evidence from the saved log to attribute it or abstain. Figure~\ref{fig:framework} illustrates our AgentAnchor framework. The primary settings described below were fixed before the held-out evaluation.


\subsection{Embedding}
\label{sec:embedding}


\paragraph{Action distribution.}
At step $t$, we build a prompt $c_t$ from the goal, the current observation, the last eight steps as plain text (each observation with its selected action), the admissible set $A_t$, and a short ReAct-style~\cite{yao2022react} planning note, one reasoning sentence, and one subgoal, written by the same model. We score each $a \in A_t$ using its length-normalized token log-likelihood under the agent model $\theta$:
\begin{equation}
P_t(a)=
\frac{
\exp\!\left(\ell_{\theta}(a\mid c_t)/|a|\right)
}{
\sum_{a'\in A_t}
\exp\!\left(\ell_{\theta}(a'\mid c_t)/|a'|\right)
},
\label{eq:pt}
\end{equation}
where $\ell_{\theta}(a\mid c_t)$ is the sum of the log-probabilities that the model assigns to the tokens (subword units) of action $a$ given $c_t$, and $|a|$ is the number of these tokens. Thus, actions that the model finds more likely after reading $c_t$ receive more mass, and dividing by $|a|$ stops long actions from losing mass only because they have more tokens. Eq.~(\ref{eq:pt}) defines the agent's policy over the finite set $A_t$ in our experiments, and we do not claim that it matches the LLM's true choice probabilities.

We quantize this distribution by largest remainder rounding to integer counts that sum to $N_q=10^6$, so each probability has a resolution of $10^{-6}$. The agent logs this fixed-point distribution, and both the embedding agent and the verifier use exactly these values. The construction of $P_t$ requires local access to model token probabilities, whereas verification does not. Because the planning note is part of $c_t$, it affects $P_t$, but it is neither logged nor needed by the verifier.


\paragraph{Session-keyed semantic synchronization.}
We encode each observation with all-MiniLM-L6-v2, a small public sentence encoder~\cite{reimers2019sentence}\cite{wang2020minilm} and assign it to its nearest semantic centroid. Let $\phi(o)\in\{0,\ldots,K-1\}$ denote this cluster label. We use $K=4$ centroids fitted once by $k$-means on 500 observations from non-evaluation scenes. The fitted centroids form a public codebook with version $\nu$. We fixed this geometry on development data before evaluation. In a development sweep over $K \in \{4, 8, 16, 32\}$, $K=4$ and $K=8$ both kept the cluster label of 56 of 58 validated moderate rewrites (96.6\%), against 82.8\% for $K=16$ and 93.1\% for $K=32$. In Appendix~\ref{app:params}, we report the full sweep.

For the current observation and the previous $w=2$ observations, where $w$ is the window length, we form the semantic window
$\Psi_t=(\phi(o_{t-2}),\phi(o_{t-1}),\phi(o_t))$, using a fixed bootstrap symbol when history is unavailable. We then derive the step key
\begin{equation}
k_t =
\mathrm{HMAC}_{\kappa}
\!\left(
D \,\|\, \mathit{sid} \,\|\, \nu \,\|\, \Psi_t
\right),
\label{eq:key}
\end{equation}
where $D$ is a fixed domain tag, and HMAC uses SHA-256.

The step key $k_t$ therefore does not directly depend on the record index, log layout, action text, or raw observation string. It depends on the observation through its semantic cluster sequence. Rewording preserves the same key only when the corresponding semantic labels remain unchanged, which held for 90.6\% of step keys under moderate paraphrase (Appendix \ref{app:ablation}). Because the fresh session identifier is also part of the HMAC input, distinct sessions use distinct key domains. Under the standard pseudorandom-function (PRF) assumption for HMAC~\cite{bellare2015new}, outputs for fresh session inputs are computationally indistinguishable from independent uniformly random values, even when the same semantic window recurs.


\paragraph{Coding row and payload bit.}
The semantic window $\Psi_t$ takes only $K^{w+1}=64$ values after the first $w$ steps, so many steps share one window and would repeat the same coding row. We therefore derive the coding row from a second public state, the most probable actions under $P_t$ (Eq.~(\ref{eq:pt})). We first map each action to a canonical label using fixed environment-specific rules, such as lowercasing and removing punctuation and extra spaces. Let the positive masses of $P_t$ be sorted as $q_{(1)}\ge q_{(2)}\ge\cdots$, and let $\mathcal{T}_t$ be the set of canonical labels of the three most probable actions. If $P_t$ has at most three positive actions, the step passes. Otherwise, we require the coding-state guard
\begin{equation*}
\frac{q_{(3)}-q_{(4)}}{q_{(3)}}\ge\gamma,\qquad \gamma=0.05,
\end{equation*}
which skips steps whose third and fourth actions have nearly equal probability, where a tiny change could swap them. We fixed $\gamma$ in an earlier pilot. On development data, it kept 77\% of steps, and no one-quantum probability change altered $\mathcal{T}_t$. We hash $\mathcal{T}_t$ together with the public coding settings (the coding-scheme tag, the top-3 rule, and $\gamma$) using SHA-256 to obtain a digest $d_t$. Small probability changes that leave $\mathcal{T}_t$ unchanged also leave $d_t$ unchanged.

We derive a nonzero $L$-bit coding row $r_t$ and the payload bit $y_t$ that step $t$ carries as
\begin{equation}
\begin{aligned}
r_t &=
\operatorname{LSB}_{L}
\!\left(
\mathrm{HMAC}_{k_t}
(D_r \,\|\, d_t \,\|\, u_t)
\right),\\
y_t &= \langle r_t,m\rangle \bmod 2 ,
\end{aligned}
\label{eq:row}
\end{equation}
where $\operatorname{LSB}_L$ keeps the $L$ least significant bits, $D_r$ is a fixed public tag that names the coding scheme, and $u_t$ is a retry counter incremented only if the truncated row is zero. We treat $r_t$ as a vector in $\mathrm{GF}(2)^L$, so $y_t$ is the parity of the payload bits that $r_t$ selects.


\paragraph{Keyed action partition and KL-bounded tilt.}
For each eligible step, we build $R=8$ candidate keyed sign functions. For candidate $j$,
\[
\pi_j(a)
=
2\,\mathrm{bit}_j
\!\left(
\mathrm{HMAC}_{k_t}(\texttt{sign}\,\|\,\bar a)
\right)-1
\in\{-1,+1\},
\]
where $\bar a$ is the canonical label of $a$. Each $\pi_j$ splits $A_t$ into a $+1$ group and a $-1$ group, which we call a keyed partition. We choose the candidate whose two groups have the most balanced probability mass under $P_t$, that is, whose signed mass is closest to zero:
\[
j^\star
=
\arg\min_j
\left|
\sum_{a\in A_t}P_t(a)\pi_j(a)
\right|,
\qquad
\pi_t=\pi_{j^\star}.
\]
We fixed $R=8$ in development before evaluation. The mean per-step evidence at $\eta=0.4$ stays within 6\% of its $R=8$ value for $R$ from 4 to 32 (Appendix~\ref{app:params}), so the choice of $R$ has little effect.

For $\sigma\in\{-1,+1\}$, we define
\begin{equation}
\widetilde P_t^{\sigma}(a)=
\frac{
P_t(a)\exp\!\left(\sigma\lambda_t\pi_t(a)\right)
}{
\sum_{a'\in A_t}
P_t(a')\exp\!\left(\sigma\lambda_t\pi_t(a')\right)
}.
\label{eq:tilt}
\end{equation}
In words, $\widetilde P_t^{+}$ ($\sigma=+1$) scales the probability of each $+1$ action by $e^{\lambda_t}$ and each $-1$ action by $e^{-\lambda_t}$ and renormalizes, while $\widetilde P_t^{-}$ ($\sigma=-1$) does the opposite. The tilt strength $\lambda_t$ sets how much mass moves. We choose the largest $\lambda_t\ge0$ such that
\[
\max_{\sigma\in\{-1,+1\}}
\mathrm{KL}
\!\left(
\widetilde P_t^{\sigma}\,\|\,P_t
\right)
\le\eta ,
\]
with $\eta=0.4$ nats (natural-log units), the largest value of our preregistered sweep (Appendix~\ref{app:params}), and a numerical cap $\lambda_t\le25$ that never binds at this $\eta$.

If $y_t=1$, we sample from $\widetilde P_t^{+}$. Otherwise, we sample from $\widetilde P_t^{-}$. Exponential tilting preserves the support of $P_t$, so the watermark cannot select an action to which the unwatermarked policy $P_t$ assigns zero probability. A step carries watermark evidence only when the logged $P_t$ is valid, at least two actions have positive support, the coding-state guard passes, and the selected keyed partition provides nonzero evidence. Otherwise, we sample directly from $P_t$. The coding row, parity, keyed partition, and $\lambda_t$ are not stored in the trajectory and are reconstructed during verification.


\paragraph{Effect of session binding.}
The keyed partition and coding row are both derived from $k_t$. Therefore, under the PRF assumption (Section~\ref{sec:embedding}), previously observed sessions do not reveal the keyed row or partition of a fresh session, even when the same $\Psi_t$ and $\mathcal{T}_t$ recur.


\subsection{Verification}
\label{sec:verification}

\paragraph{Recomputation.}
Given a trajectory log and the secret key $\kappa$, we first check that all records contain one consistent, syntactically valid session identifier. Missing, malformed, or inconsistent session context produces no usable evidence and causes abstention. The verifier then recomputes the semantic labels and windows, step keys from Eq.~\eqref{eq:key}, coding-state digests and rows from Eq.~\eqref{eq:row}, and the keyed tilted distributions from Eq.~\eqref{eq:tilt}. All required inputs other than $\kappa$ come from the structured log or the fixed public configuration. Record indices are not used in the key or coding row, although the received record order determines the semantic windows.


\paragraph{Likelihood evidence.}
Let $E$ denote the usable eligible steps after exact duplicate records, such as copies inserted by an attacker, are removed. For each $t\in E$, we compute
\[
\Delta_t=
\log\widetilde P_t^{+}(a_t)
-
\log\widetilde P_t^{-}(a_t),
\]
\[
b_t=
\tfrac{1}{2}
\big[
\log\widetilde P_t^{+}(a_t)
+
\log\widetilde P_t^{-}(a_t)
\big].
\]
$\Delta_t$ measures how much more likely the selected action $a_t$ is under $\widetilde P_t^{+}$ than under $\widetilde P_t^{-}$, so its sign points to the bit that the step carries. $b_t$ is the average of the two log-likelihoods and is the same for every candidate. For a candidate identifier $m'$, let
$y_t(m')=\langle r_t,m'\rangle\bmod2$ and
$\sigma_t(m')=2y_t(m')-1$, which is the tilt direction $\sigma$ of Eq.~\eqref{eq:tilt} that $m'$ implies at step $t$. Its score is
\begin{equation}
\begin{aligned}
S_{\kappa}(m')
&=
\sum_{t\in E}
\left[
b_t+\frac{1}{2}\sigma_t(m')\Delta_t
\right] \\
&=
\sum_{t\in E}
\log
\widetilde P_t^{\sigma_t(m')}(a_t).
\end{aligned}
\label{eq:score}
\end{equation}
Thus, $S_{\kappa}(m')$ is the conditional log-likelihood of the observed eligible actions under key $\kappa$ and candidate identifier $m'$. We decode
\[
\hat m=
\arg\max_{m'\in\{0,1\}^{L}}
S_{\kappa}(m').
\]
For claimed-owner detection, the identifier $m_0$ is fixed before verification, and we use $S_{\kappa}(m_0)$ directly. This keeps claimed-owner detection separate from open-set owner attribution.


\paragraph{Wrong-key calibration.}
The score distribution has no simple null form because candidate identifiers share coding rows and each step depends on its logged distribution. So, we compare the scores under $\kappa$ with those under $M$ random keys $\kappa_1,\ldots,\kappa_M$, which the verifier draws independently of the log. Each random key goes through exactly the same recomputation and scoring. We compute
\begin{equation}
\begin{aligned}
p_{\mathrm{det}}
&=
\frac{
1+
\left|
\left\{
i:
S_{\kappa_i}(m_0)
\ge
S_{\kappa}(m_0)
\right\}
\right|
}{
M+1
},\\
p_{\mathrm{id}}
&=
\frac{
1+
\left|
\left\{
i:
\max_{m'}S_{\kappa_i}(m')
\ge
\max_{m'}S_{\kappa}(m')
\right\}
\right|
}{
M+1
}.
\end{aligned}
\label{eq:pvalues}
\end{equation}
$p_{\mathrm{det}}$ is the $p$-value for claimed-owner detection, and $p_{\mathrm{id}}$ is the $p$-value for open-set attribution using the best score over all identifiers. A small value means that the evidence under $\kappa$ beats almost all random keys.

When the tested log carries no information about $\kappa$, $\kappa$ is exchangeable with the random keys. The added one in Eq.~\eqref{eq:pvalues}, which counts $\kappa$ itself among the $M+1$ keys, then gives $\Pr[p\le\alpha]\le\alpha$~\cite{phipson2016permutation}, so such a log passes at significance level $\alpha$ with probability at most $\alpha$. In Appendix~\ref{app:forgery}, we show that this bound also holds, up to a negligible term, for a forger trained only on other sessions. Session reuse is not covered, so we also evaluate learned forgery directly.

The maximization over all $2^L=256$ candidate identifiers is repeated under every random key. Hence, open-set search is included in the empirical null rather than treated as a collection of independent hypotheses. We use $\alpha=10^{-3}$, with $M=1999$ for clean final verification and $M=999$ in attack evaluation. Wrong-key calibration follows the general approach used by SeqWM~\cite{an2026sequential}.


\paragraph{Decision.}
For each usable step, the verifier also computes a public coding identity, a hash of the coding-scheme tag $D_r$, the key design, the codebook version $\nu$, $\mathit{sid}$, $\Psi_t$, and $d_t$. Steps with the same identity have the same coding row. Let $C$ be the number of distinct identities in the verification unit, and let $\rho$ be the GF(2) rank of its coding rows.

We attribute the verification unit to $\hat m$ only when
$p_{\mathrm{id}}\le\alpha$, $C\ge L$, $\rho=L$, and the maximum in Eq.~\eqref{eq:score} is unique. Rank below $L$ leaves a nontrivial kernel in the row matrix and therefore makes several identifiers $m'$ structurally indistinguishable. Even at full rank, a score tie causes abstention. If unique attribution is unavailable but $p_{\mathrm{det}}\le\alpha$ for a predeclared $m_0$, we report claimed-owner detection. Otherwise, we abstain.


\paragraph{Pooling.}
For several trajectories claimed to share one owner identifier, we reconstruct each trajectory independently and concatenate their usable per-step evidence. We then compute the score of Eq.~\eqref{eq:score} and the $p$-values of Eq.~\eqref{eq:pvalues} on the combined evidence while computing $C$ and $\rho$ over the union of coding identities and rows. The same random key is used across every trajectory in a pooled verification unit. Pooling can therefore add coding diversity and statistical evidence when a single trajectory is insufficient. Our primary pooled evaluation uses disjoint groups of five trajectories.



\begin{table*}[t]
\caption{Clean verification ($M=1999$, $\alpha=10^{-3}$). $n$: verification units. Coded steps: share of logged steps carrying a coded bit. $\bar{C}$: mean distinct coding identities per unit. Rank-$L$: units whose coding rows reach full GF(2) rank. Det.: claimed-owner detections. Exact: $\hat{m}$ equals the owner identifier and passes the $C\ge L$, $\rho=L$, and uniqueness checks before the $p_{\mathrm{id}}$ test. Correct/Wrong: accepted attributions to the owner or another identifier. FD/FA: false detections and attributions on unwatermarked controls of the same tasks.}
\label{tab:clean}
\centering
\scriptsize
\setlength{\tabcolsep}{4.2pt}
\begin{tabular}{llrrrrrrrrrr}
\toprule
& & & \multicolumn{3}{c}{Evidence} &
\multicolumn{4}{c}{Watermarked} &
\multicolumn{2}{c}{Unwatermarked} \\
\cmidrule(lr){4-6}\cmidrule(lr){7-10}\cmidrule(lr){11-12}
Benchmark & Unit & $n$ &
Coded steps (\%) $\uparrow$ &
$\bar C$ $\uparrow$ &
Rank-$L$ $\uparrow$ &
Det. $\uparrow$ &
Exact $\uparrow$ &
Correct $\uparrow$ &
Wrong $\downarrow$ &
FD $\downarrow$ &
FA $\downarrow$ \\
\midrule
\multirow{2}{*}{ALFWorld}
 & Single$^\dagger$ & 50 & 75.4 & 34.2 & 49/50 & 47/50 & 48/50 & 42/50 & 0/50 & 0/50 & 0/50 \\
 & Pool of five    & 10 & 75.4 & 170.9 & 10/10 & 10/10 & 10/10 & 10/10 & 0/10 & 0/10 & 0/10 \\
\midrule
\multirow{2}{*}{WebShop-1K}
 & Single$^\dagger$ & 30 & 58.9 & 18.8 & 18/30 & 17/30 & 17/30 & 13/30 & 0/30 & 0/30 & 0/30 \\
 & Pool of five    & 6  & 58.9 & 94.2 & 6/6 & 6/6 & 6/6 & 6/6 & 0/6 & 0/6 & 0/6 \\
\bottomrule
\end{tabular}

\smallskip
\begin{minipage}{\textwidth}\scriptsize
$^\dagger$Single-trajectory attribution is descriptive. On ALFWorld, its development-stage identification score (0.676) did not meet our preregistered 0.80 gate. WebShop-1K transfers the frozen ALFWorld operating point and has no separate single-trajectory identification gate.
\end{minipage}
\end{table*}


\begin{table*}[t]
\caption{Learned forgery (a) and log transformations (b), all at $M=999$ and $\alpha=10^{-3}$, where the smallest attainable $p$ equals $\alpha$. Header numbers are unit counts. (a) The attacker learns from $N$ owner-watermarked logs and rewrites unwatermarked targets. Overlap: target sessions seen during learning. FD/FA: forged units passing claimed-owner detection or owner attribution, worse than the weak and strong attackers. Pooled groups reuse trajectories across three fixed partitions. (b) Det.: detections. Correct (wrong): attributions to the owner (to another identifier). Pool: correctly attributed pools of five. $^{*}$: unadjusted exact McNemar $p<0.05$ against the clean row.}
\label{tab:forgery}
\centering
\scriptsize

\textbf{(a) Learned forgery: FD / FA $\downarrow$}

\smallskip
\setlength{\tabcolsep}{4.3pt}
\begin{tabular}{lrrrrrrrr}
\toprule
& \multicolumn{4}{c}{ALFWorld} & \multicolumn{4}{c}{WebShop-1K} \\
\cmidrule(lr){2-5}\cmidrule(lr){6-9}
Sessions &
$N$ & Overlap & Single (50) & Pooled (30) &
$N$ & Overlap & Single (30) & Pooled (18) \\
\midrule
Fresh  & 10  & 0  & 0/0  & 0/0  & 10 & 0  & 0/0 & 0/0 \\
Fresh  & 50  & 0  & 0/0  & 0/0  & 35 & 0  & 0/0 & 0/0 \\
Fresh  & 200 & 0  & 1$^\ddagger$/0 & 0/0 & \multicolumn{4}{c}{n/r} \\
Reused & 10  & 8  & 1/0  & 1/0  & 10 & 10 & 0/0 & 1/0 \\
Reused & 50  & 32 & 1/0  & 2/0  & 35 & 30 & 1/0 & 4/0 \\
Reused & 200 & 50 & 44/25 & 30/30 & \multicolumn{4}{c}{n/r} \\
\bottomrule
\end{tabular}

\medskip
\textbf{(b) Verification after log transformation}

\smallskip
\setlength{\tabcolsep}{3.6pt}
\begin{tabular}{llrrrrrr}
\toprule
& & \multicolumn{3}{c}{ALFWorld} & \multicolumn{3}{c}{WebShop-1K} \\
\cmidrule(lr){3-5}\cmidrule(lr){6-8}
Transformation & Family &
Det. (50) & Correct (wrong) & Pool (10) &
Det. (30) & Correct (wrong) & Pool (6) \\
\midrule
None (clean) & -- &
46 & 37 (0) & 10 &
18 & 13 (0) & 6 \\

Step renumbering & Restructuring &
47 & 37 (0) & 10 &
19 & 12 (0) & 6 \\

Interface padding & Restructuring &
47 & 38 (0) & 10 &
\multicolumn{3}{c}{n/r} \\

$P_t$ shift ($10^{-6}$) & Logging noise &
47 & 37 (0) & 10 &
17 & 14 (0) & 6 \\

$P_t$ relative noise (1\%) & Logging noise &
44 & 30 (0) & 10 &
15 & 13 (0) & 6 \\

Moderate paraphrase & Rewriting &
44 & 23$^\ast$ (1) & 10 &
16 & 12 (0) & 6 \\

Heavy paraphrase & Rewriting &
43 & 28$^\ast$ (1) & 10 &
\multicolumn{3}{c}{n/r} \\

EN$\rightarrow$ES$\rightarrow$EN & Rewriting &
45 & 35 (0) & 10 &
17$^\S$ & 13 (0) & 6 \\

Adaptive paraphrase & Rewriting &
29$^\ast$ & 17$^\ast$ (0) & 10 &
\multicolumn{3}{c}{n/r} \\

Step erasure (30\%) & Structural &
13$^\ast$ & 0$^\ast$ (0) & 8 &
10$^\ast$ & 5$^\ast$ (2) & 5 \\

Truncation (keep 50\%) & Structural &
33$^\ast$ & 4$^\ast$ (0) & 10 &
13 & 6$^\ast$ (0) & 5 \\

Step insertion (20\%) & Structural &
32$^\ast$ & 11$^\ast$ (0) & 10 &
15 & 9 (0) & 6 \\

Random reordering & Structural &
3$^\ast$ & 0$^\ast$ (0) & 1 &
5$^\ast$ & 2$^\ast$ (0) & 4 \\

Action substitution (30\%) & Structural &
28$^\ast$ & 4$^\ast$ (0) & 10 &
8$^\ast$ & 3$^\ast$ (0) & 4 \\

Action renaming & Evidence editing &
0$^\ast$ & 0$^\ast$ (0) & 0 &
\multicolumn{3}{c}{n/r} \\

Digest-boundary $P_t$ tamper & Evidence editing &
0$^\ast$ & 0$^\ast$ (0) & 0 &
1$^\ast$ & 0$^\ast$ (0) & 0 \\

Session ID stripped & Evidence editing &
0$^\ast$ & 0$^\ast$ (0) & 0 &
0$^\ast$ & 0$^\ast$ (0) & 0 \\

Session ID replaced & Evidence editing &
0$^\ast$ & 0$^\ast$ (0) & 0 &
0$^\ast$ & 0$^\ast$ (1) & 0 \\
\bottomrule
\end{tabular}

\smallskip
\begin{minipage}{\textwidth}\scriptsize
$^\ddagger$Our preregistered ALFWorld forgery criterion required zero events and therefore failed because of this single false detection, although no fresh-session false attribution occurred. 
$^\S$The WebShop-1K round-trip attack was regenerated after the frozen version produced identity outputs and is therefore exploratory. 
n/r: not run.
\end{minipage}
\end{table*}


\begin{figure*}[t]
  \centering
  \begin{subfigure}[b]{0.32\textwidth}
    \includegraphics[width=\linewidth]{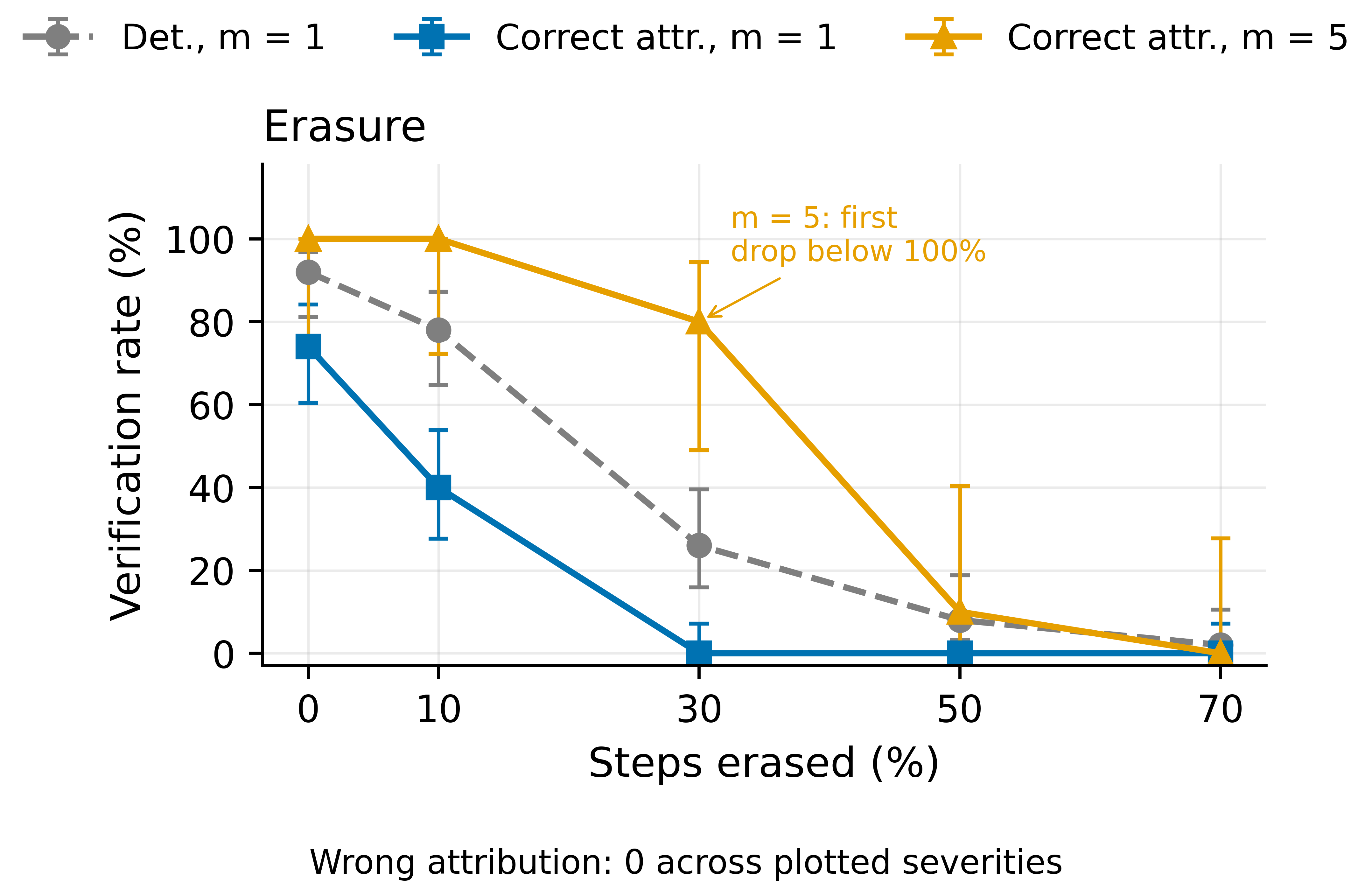}
    \caption{Step erasure}
    \label{fig:rob-era}
  \end{subfigure}\hfill
  \begin{subfigure}[b]{0.32\textwidth}
    \includegraphics[width=\linewidth]{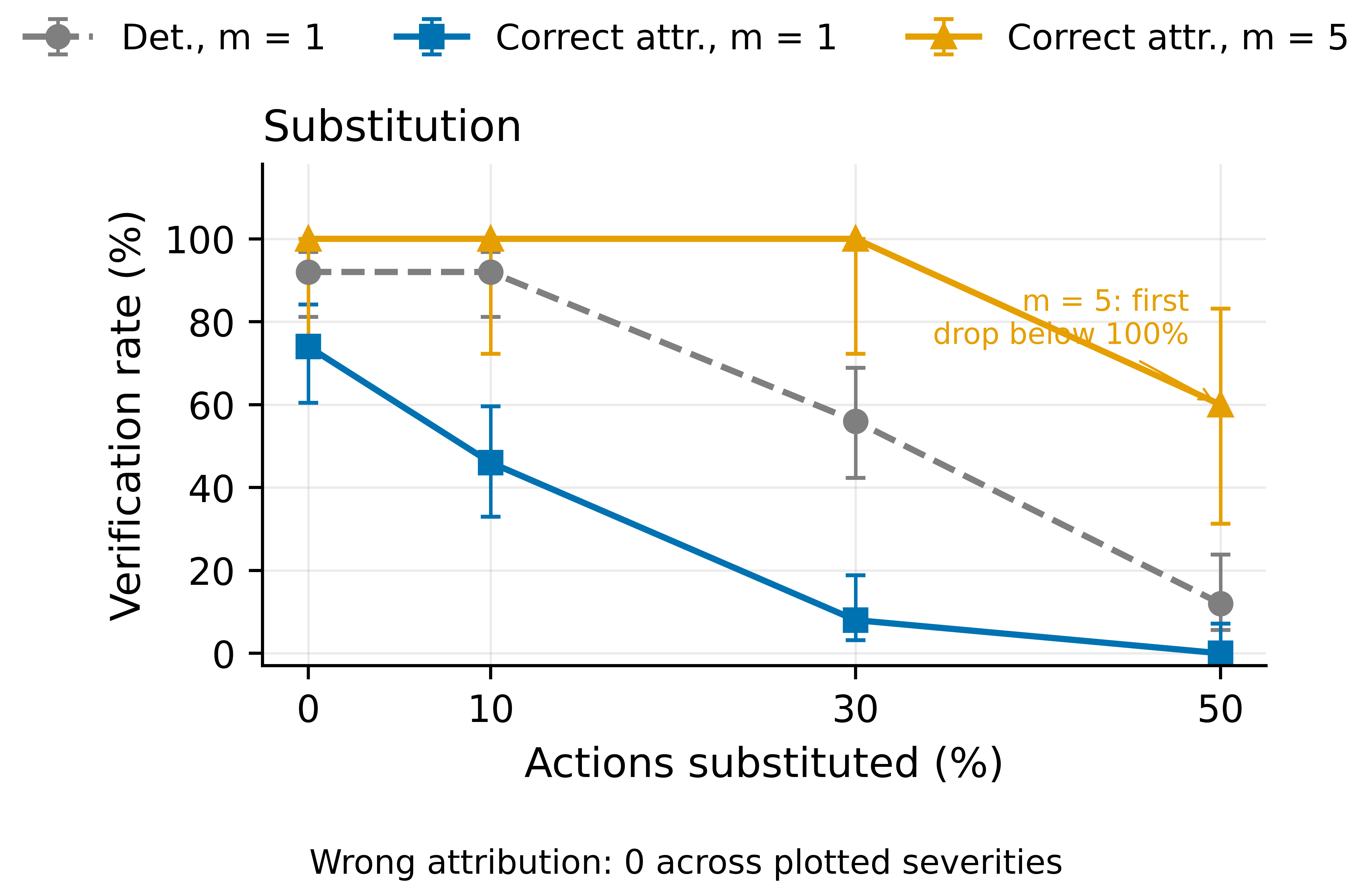}
    \caption{Action substitution}
    \label{fig:rob-sub}
  \end{subfigure}\hfill
  \begin{subfigure}[b]{0.32\textwidth}
    \includegraphics[width=\linewidth]{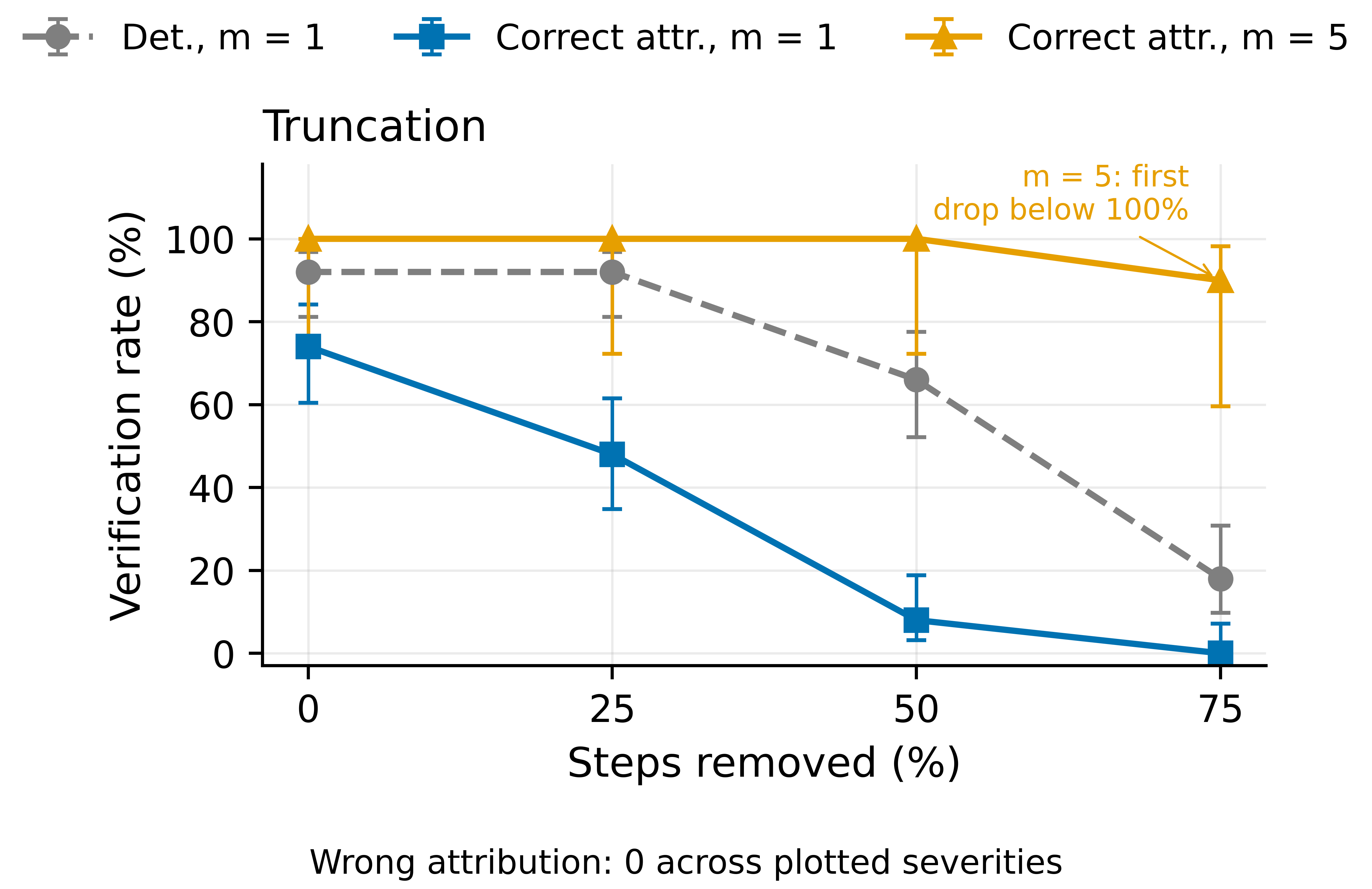}
    \caption{Truncation}
    \label{fig:rob-tru}
  \end{subfigure}
  \caption{Robustness degradation under structural edits on ALFWorld. Each edit is applied at increasing severity to the 50 watermarked evaluation trajectories. Curves show single-trajectory detection, single-trajectory correct attribution, and correct attribution of the 10 primary pools of five, with 95\% Wilson intervals ($M=999$, $\alpha=10^{-3}$). For truncation, severity is the share of steps removed from the end. No wrong attribution occurs at any plotted severity.}
  \label{fig:robustness}
\end{figure*}


\section{Experiments and Analysis}
\label{sec:experiments}

We evaluate AgentAnchor on ALFWorld~\cite{shridhar2020alfworld} and WebShop-1K~\cite{yao2022webshop} using the same frozen watermark core and pinned \texttt{gemma-2-9b-it} model. ALFWorld uses 50 held-out \texttt{valid\_unseen} tasks across six task types with a 50-step cap. All design choices use only \texttt{valid\_seen} data and were checked on 25 fresh confirmation tasks. WebShop-1K uses the official 1,000-product subset with its default attribute-and-option goals, since only 13 human-written instructions target this catalogue. It has 30 frozen goals with distinct targets, a 100-step cap, and a $K=4$ codebook fitted on 500 non-evaluation observations. We transfer the ALFWorld operating point without retuning and use conservative action canonicalization. Search queries are generated greedily without the key, so search steps carry no watermark bit. Paraphrases modify at most 25 observations per trajectory and must pass bidirectional natural language inference (NLI) and protected-token checks. For learned forgery, attackers observe $N$ logs re-embedded under the owner's key from base pools of 75 ALFWorld and 35 WebShop-1K logs, then modify unwatermarked targets under fresh or reused sessions. The weak attacker learns action preferences per semantic context, and the strong attacker also conditions on the coding digest. We test $N\in\{10,50,200\}$ and $N\in\{10,35\}$, respectively. WebShop-1K omits several attacks by predeclared protocol.


\subsection{Clean Verification and Evidence Sufficiency}
\label{sec:exp-clean}

Our Table~\ref{tab:clean} supports log-only verification and calibrated abstention that no unwatermarked unit was detected or attributed, and no watermarked unit was attributed to a wrong identifier. Pooling uses the same verifier and mainly resolves evidence insufficiency. The coding digest adds row diversity beyond semantic windows, since among embedded steps, an ALFWorld trajectory visits 14.6 distinct semantic windows on average but yields 34.2 distinct usable coding states. On WebShop-1K, single-action search steps and unstable coding states leave 12 of the 17 unattributed trajectories without enough coding states or rank. We therefore treat five-trajectory pooling as the primary configuration and single-trajectory attribution as descriptive. Every embedded step met the KL budget, supporting the bounded-distortion goal.


\subsection{Learned Forgery and Session Freshness}
\label{sec:exp-forgery}

Our Table~\ref{tab:forgery}(a) isolates session binding, since both conditions share attackers, base logs, and targets. Under fresh sessions, no forgery was attributed to the owner on either benchmark. One ALFWorld false detection fails our zero-event criterion but is consistent with Appendix~\ref{app:forgery}. At $N=200$, the strong attacker changed 84\% of target actions, yet the median $p_{\mathrm{det}}$ of its forgeries was 0.57, close to the 0.5 expected without owner evidence. Smaller $N$ changed few actions. Reused sessions let the same attack produce false attributions once each session was observed repeatedly, which WebShop-1K's small pool cannot reach. Session freshness is therefore a necessary deployment assumption, and the results show empirical resistance to the tested attackers, not unforgeability.


\subsection{Robustness Under Log Transformations}
\label{sec:exp-robust}

We show in our Table~\ref{tab:forgery}(b) and Figure~\ref{fig:robustness} that robustness follows the synchronization and evidence mechanisms. Renumbering, interface padding, and small probability edits preserve pooled attribution because they leave the keyed state unchanged, and the guarded digest never changes under them. Rewriting lowers single-trajectory attribution when it changes cluster labels (Table~\ref{tab:ablation}), yet pooling recovers every ALFWorld group. WebShop-1K rewrites changed no label, so they do not test semantic robustness. Pooling compensates for moderate evidence loss, but severe structural edits and direct evidence edits mark clear failure boundaries. Wrong attribution is rare but not absent: eight of 2,970 edited single-trajectory decisions, all at the smallest attainable $p$-value, and one WebShop-1K sensitivity-partition pool.


\subsection{Discussion and Limitations}
\label{sec:limitations}

Single trajectories are often insufficient, especially on WebShop-1K, and single-trajectory identification missed its preregistered development target (0.676 versus 0.80). Embedding requires local access to candidate-action likelihoods, and we test only one such model. Verification assumes a private key and trustworthy logged probabilities. AgentAnchor does not authenticate logs or prevent replay, so session freshness must be provided externally. Our forgery bound holds only per decision, for fresh sessions, and under the PRF assumption (Appendix~\ref{app:forgery}). One fresh-session false detection failed our zero-event criterion, and session reuse enabled false attribution. Finally, we make no utility-parity claim. The per-step KL budget does not bound effects that compound over closed-loop steps.



\section{Conclusion}
\label{sec:conclusion}

An LLM agent watermark should be judged not only by what a verifier detects but also by what observers can learn from its logs. Behavioral watermarking for LLM agents must therefore provide recoverable evidence and limit cross-session transfer. In our experiments, pooled verification attributed owners correctly under several log transformations, while severe corruption removed the evidence and session reuse let forgeries be attributed to the owner. These results position our AgentAnchor as a bounded attribution mechanism that works when session identifiers are fresh and enough coded steps are pooled. Future work should study stronger adaptive forgers, authenticated session issuance, and larger real-world Web deployments, and evaluate task utility with stronger base agents.


\bibliographystyle{ACM-Reference-Format}
\bibliography{AgentAnchor}

\appendix



\section*{Use of Generative AI}

We used generative AI tools (Claude Opus 5.5 and GPT Sol 5.6) in this work only for limited language assistance, including grammar correction, improving readability, and formatting in Overleaf. These tools were not used to develop the core research ideas, formulate hypotheses, or design the methodology. We also used AI assistance to discuss how to present the benchmarks, evaluation protocol, and results more clearly. All AI-assisted text and literature suggestions were reviewed and verified by the authors, and relevant references were checked against their original sources. The authors take full responsibility for the final content of this work, including all claims, results, and conclusions.


\section*{Appendix}


\section{Parameter Selection and Ablation Study}
\label{app:params}\label{app:ablation}

\begin{table*}[t]

\caption{An ablation study of our AgentAnchor framework on ALFWorld. $\star$ marks the frozen setting. Parts (a), (b), and (d) replay the evaluation logs and were not used to select any parameter. (a) Each key design re-embeds the owner identifier into the logged $P_t$ of the 50 watermarked evaluation logs with the same channel, coding, and decoder. Sync is the share of steps whose key, recomputed from the edited log, equals the key used at embedding. Names in parentheses mark the prior work whose keying a baseline approximates. Padding is the interface padding of Table~\ref{tab:forgery}(b). Det.\ counts claimed-owner detections out of 50 ($M=999$, $\alpha=10^{-3}$) and is not directly comparable with the live results of Table~\ref{tab:forgery}(b). Only the frozen design's resistance to learned forgery was measured (Table~\ref{tab:forgery}(a)). (b) Eligible is the share of steps that carry a bit. Hard channels only require their split to exist, whereas the soft tilt applies the full eligibility rule, which we mentioned in Section 3. TV is the total variation distance from $P_t$ to the embedding distribution, which is $\widetilde P_t^{+}$ for the soft tilt. $\bar J$ is the mean over eligible steps of $\tfrac12[\mathrm{KL}(\widetilde P_t^{+}\|\widetilde P_t^{-})+\mathrm{KL}(\widetilde P_t^{-}\|\widetilde P_t^{+})]$. The preregistered sweep is $\eta\in\{0.05,0.1,0.2,0.4\}$, $\eta=0.6$ is an ablation, and $\eta=0.8$, where the cap $\lambda_t\le 25$ binds, is omitted. (c) Correct unique identification on the 25 development confirmation tasks, with two embedding seeds, two development keys, and the seed-101 partition, giving 100, 48, 32, and 20 decisions for pools of 1, 2, 3, and 5. These screening tests use $M=199$ and $\alpha=5\times10^{-3}$, so they are not comparable with the 0.676 of Table~\ref{tab:clean}. (d) A share of the coded steps of each trajectory takes its coding row and evidence from a wrong key. Counts are taken before the $p$-value test. Gaussian elimination treats the sign of each $\Delta_t$ as a hard parity bit.}
\label{tab:ablation}

\centering
\footnotesize
\setlength{\tabcolsep}{3pt}

\begin{tabular}{llcccccccc}
\toprule
\multicolumn{10}{l}{\textbf{(a) Key synchronization and detection (counterfactual replay, 50 evaluation logs)}} \\
\midrule
& & \multicolumn{6}{c}{Sync $\uparrow$} & \multicolumn{2}{c}{Det.\ $\uparrow$} \\
\cmidrule(lr){3-8}\cmidrule(lr){9-10}
Key design & Key input & Renumber & Padding & Para.\ mod. & Para.\ adapt. & Erase 30\% & No session & Para.\ adapt. & Erase 30\% \\
\midrule
Session $+$ window $\star$ & $\mathit{sid}$, $\Psi_t$ & 1.000 & 1.000 & 0.906 & 0.761 & 0.677 & 0.000 & 41 & 18 \\
Window only & $\Psi_t$ & 1.000 & 1.000 & 0.906 & 0.761 & 0.672 & 1.000 & 37 & 13 \\
Goal $+$ window & goal cluster, $\Psi_t$ & 1.000 & 1.000 & 0.906 & 0.761 & 0.670 & 1.000 & 33 & 19 \\
Action window (SeqWM) & last three actions & 1.000 & 1.000 & 1.000 & 1.000 & 0.360 & 1.000 & 48 & 2 \\
Previous action (TRACE) & last action & 1.000 & 1.000 & 1.000 & 1.000 & 0.705 & 1.000 & 47 & 14 \\
Positional (Agent Guide) & logged index & 0.000 & 1.000 & 1.000 & 1.000 & 1.000 & 1.000 & 47 & 43 \\
Index $+$ observation & index, observation & 0.000 & 1.000 & 0.474 & 0.472 & 1.000 & 1.000 & 9 & 42 \\
Index $+$ action set & index, $A_t$ signature & 0.000 & 0.000 & 1.000 & 1.000 & 1.000 & 1.000 & 48 & 41 \\
\midrule

\multicolumn{10}{l}{\textbf{(b) Embedding channel (2,375 logged evaluation steps)}} \\
\midrule
Channel & Setting & Eligible (\%) & KL mean $\downarrow$ & KL max $\downarrow$ & TV mean $\downarrow$ & $\bar J$ $\uparrow$ & & & \\
\midrule
Keyed half-mass split (hard) & -- & 98.2 & 0.814 & 12.02 & 0.500 & -- & & & \\
Median-bin split (hard) & -- & 87.9 & 1.135 & 3.19 & 0.608 & -- & & & \\
Soft tilt & $\eta=0.05$, $R=8$, label & 75.4 & 0.050 & 0.050 & 0.147 & 0.19 & & & \\
Soft tilt & $\eta=0.1$, $R=8$, label & 75.4 & 0.100 & 0.100 & 0.205 & 0.38 & & & \\
Soft tilt & $\eta=0.2$, $R=8$, label & 75.4 & 0.200 & 0.200 & 0.282 & 0.76 & & & \\
Soft tilt $\star$ & $\eta=0.4$, $R=8$, label & 75.4 & 0.400 & 0.400 & 0.380 & 1.62 & & & \\
Soft tilt & $\eta=0.6$, $R=8$, label & 75.4 & 0.600 & 0.600 & 0.445 & 2.81 & & & \\
Soft tilt & $\eta=0.4$, $R=8$, rank & 55.6 & 0.400 & 0.400 & 0.383 & 1.62 & & & \\
\midrule

\multicolumn{10}{l}{\textbf{(c) Payload length and pool size (development data, before evaluation)}} \\
\midrule
Payload & Identifiers $2^L$ & Pool 1 & Pool 2 & Pool 3 & Pool 5 $\star$ & Det., pool 1 & & & \\
\midrule
$L=2$ & 4 & 0.85 & 1.00 & 1.00 & 1.00 & 0.89 & & & \\
$L=4$ & 16 & 0.81 & 1.00 & 1.00 & 1.00 & 0.94 & & & \\
$L=6$ & 64 & 0.80 & 1.00 & 1.00 & 1.00 & 0.91 & & & \\
$L=8$ $\star$ & 256 & 0.76 & 0.98 & 1.00 & 1.00 & 0.92 & & & \\
\midrule

\multicolumn{10}{l}{\textbf{(d) Decoder under wrong-key corruption of coded steps (single trajectories, counts out of 50)}} \\
\midrule
Decoder & Outcome & 0\% & 5\% & 10\% & 20\% & 30\% & & & \\
\midrule
Log-likelihood $\star$ & correct, unique $\uparrow$ & 48 & 48 & 47 & 42 & 30 & & & \\
Gaussian elim. & correct $\uparrow$ & 17 & 10 & 14 & 4 & 5 & & & \\
Gaussian elim. & confidently wrong $\downarrow$ & 32 & 39 & 35 & 45 & 44 & & & \\
\bottomrule
\end{tabular}

\end{table*}

\begin{table*}[t]
\centering
\footnotesize
\setlength{\tabcolsep}{3pt}

\begin{minipage}[t]{0.485\textwidth}

\caption{Semantic-geometry sweep on development data. The first four columns give the share of 58 validated moderate rewrites whose cluster label is unchanged, for the all-MiniLM-L6-v2 and all-mpnet-base-v2 encoders and codebooks fitted on 250 or 500 observations. For MiniLM with 500 observations and $w=2$, Heavy is the same share for heavy rewrites, Margin is the median Euclidean distance from an observation embedding to its nearest cluster boundary, and Windows is the mean number of distinct $\Psi_t$ per trajectory. With $K=4$, $w=0$, 1, and 2 give 2.8, 7.4, and 13.8 windows, while one deleted step changes at most $w+1$ windows.}
\label{tab:geometry}

\centering
\footnotesize
\setlength{\tabcolsep}{3pt}

\begin{tabular}{rccccccc}
\toprule
& \multicolumn{2}{c}{MiniLM $\uparrow$} & \multicolumn{2}{c}{MPNet $\uparrow$} & \multicolumn{3}{c}{MiniLM, 500, $w=2$} \\
\cmidrule(lr){2-3}\cmidrule(lr){4-5}\cmidrule(lr){6-8}
$K$ & 250 & 500 & 250 & 500 & Heavy $\uparrow$ & Margin $\uparrow$ & Windows \\
\midrule
4 $\star$ & 0.948 & 0.966 & 0.948 & 0.914 & 0.914 & 0.229 & 13.8 \\
8         & 0.931 & 0.966 & 0.914 & 0.897 & 0.914 & 0.256 & 26.9 \\
16        & 0.931 & 0.828 & 0.897 & 0.914 & 0.862 & 0.220 & 34.9 \\
32        & 0.862 & 0.931 & 0.845 & 0.879 & 0.897 & 0.217 & 36.4 \\
\bottomrule
\end{tabular}

\end{minipage}\hfill
\begin{minipage}[t]{0.485\textwidth}

\caption{Sensitivity to the number $R$ of candidate sign functions was replayed on the 2,375 logged ALFWorld evaluation steps with label binding. TV and $\bar J$ follow Table~\ref{tab:ablation}(b), and 75.4\% of steps are eligible in every cell. For small tilts, $\mathrm{KL}(\widetilde P_t^{\sigma}\|P_t)\approx\lambda_t^2\mathrm{Var}_{P_t}(\pi_t)/2$, while the per-step separation averaged in $\bar J$ is close to $2\lambda_t^2\mathrm{Var}_{P_t}(\pi_t)$, so $\bar J\approx4\eta$. The replay selects nothing, since $R=8$ was fixed before evaluation.}
\label{tab:rsweep}

\centering
\footnotesize
\setlength{\tabcolsep}{4pt}

\begin{tabular}{rccccc}
\toprule
& \multicolumn{4}{c}{$\bar J$ $\uparrow$} & TV $\downarrow$ \\
\cmidrule(lr){2-5}\cmidrule(lr){6-6}
$R$ & $\eta=0.05$ & $\eta=0.1$ & $\eta=0.2$ & $\eta=0.4$ $\star$ & $\eta=0.4$ \\
\midrule
4         & 0.19 & 0.37 & 0.74 & 1.54 & 0.370 \\
8 $\star$ & 0.19 & 0.38 & 0.76 & 1.62 & 0.380 \\
16        & 0.19 & 0.38 & 0.78 & 1.67 & 0.388 \\
32        & 0.19 & 0.39 & 0.79 & 1.71 & 0.391 \\
\bottomrule
\end{tabular}

\end{minipage}

\end{table*}

All of our parameters were fixed before evaluation, and no evaluation data were used for tuning. Our Table~\ref{tab:geometry} shows that, with MiniLM and 500 observations, $K=4$ and $K=8$ achieve the same highest moderate-rewrite stability. For that reason, we retain the frozen $K=4$ without claiming that it is optimal. Our Table~\ref{tab:rsweep} shows that $\bar J$ remains close to $4\eta$ and changes by less than 6\% relative to its value at $R=8$. Thus, $\eta$ mainly controls the per-step evidence, and we use the largest preregistered value. Our primary pooled analysis uses disjoint groups of five fixed by seed 101. Seeds 211 and 307 only regroup the same trajectories to provide overlapping sensitivity views. Pools of three already achieved full identification in development (Table~\ref{tab:ablation}(c)). GPU cost determined the 25-step paraphrase cap and the use of 30 WebShop-1K goals. WebShop-1K uses its official baseline's 100-step cap and a conservative canonicalizer because the ALFWorld rules merged distinct product options. We omit light, heavy, and adaptive paraphrases, interface padding, and renaming for WebShop-1K. After the freeze, we repaired the round-trip generator, which had returned unchanged text, and the masking of overlapping protected spans; neither repair changed the evaluation logs or verification procedure. The 2,970 edited decisions in Section~\ref{sec:exp-robust} cover 39 ALFWorld and 34 WebShop-1K conditions, including additional severities, $P_t$ mixing, swaps, validity flips, goal and session removal, and a combined attack beyond those summarized in Table~\ref{tab:forgery}(b). Table~\ref{tab:ablation}(a) shows that action-based keys are insensitive to observation rewriting but lose history under erasure, whereas index-based keys survive erasure when logged indices remain unchanged but fail under renumbering. The session-keyed design follows almost the same synchronization pattern as window-only keying while the session is present, but removing the session breaks its synchronization. The hard channels exceed the KL budget, and hard-bit Gaussian elimination is often confidently wrong even when no corruption is introduced.


\section{Fresh-Session Forgery Bound}
\label{app:forgery}

We state what session binding guarantees against a forger in this appendix. Let $X$ be the verification unit that the verifier receives, and let $G_\kappa$ be either $S_\kappa(m_0)$ or $\max_{m'}S_\kappa(m')$, computed on $X$ under key $\kappa$. The key enters $G_\kappa$ only through the step keys of Eq.~(2). All other inputs come from $X$ or the public configuration.

We assume three conditions. (A1) The keys $\kappa,\kappa_1,\dots,\kappa_M$ are independent uniform 256-bit strings, and the verifier draws $\kappa_1,\dots,\kappa_M$ independently of $X$. (A2) An efficient forger sees watermarked logs only from a set $\Omega$ of session identifiers. It may also use any data that does not depend on $\kappa$, but it sees no verifier output. (A3) Every log in $X$ carries a valid session identifier outside $\Omega$.


\begin{proposition}[Fresh-session forgery bound]
\label{prop:fresh}
Under (A1)--(A3), $\Pr[p_{\mathrm{det}}\le\alpha]\le\alpha+\epsilon$ and $\Pr[p_{\mathrm{id}}\le\alpha]\le\alpha+\epsilon$, where $\epsilon\le(M+1)\,\mathrm{Adv}^{\mathrm{prf}}_{\mathrm{HMAC}}$ and $\mathrm{Adv}^{\mathrm{prf}}_{\mathrm{HMAC}}$ is the largest advantage of an efficient distinguisher against HMAC-SHA256 as a pseudorandom function~\cite{bellare2015new}.
\end{proposition}


\begin{proof}
The tag $D$ is fixed, and the session identifier has a fixed length of 128 bits, so every HMAC input of Eq.~(2) built from $X$ differs from every input used to produce the forger's view. Firstly, we replace $\mathrm{HMAC}_\kappa$ everywhere by a uniformly random function $F$. A distinguisher can simulate the forger and the verifier and check whether $p\le\alpha$, so this step changes the probability of that event by at most one PRF advantage. The values of $F$ at the inputs of $X$ are independent of the forger's view and so of $X$. Secondly, we replace each $\mathrm{HMAC}_{\kappa_i}$ by an independent random function, which costs at most $M$ further advantages. Then $G_\kappa,G_{\kappa_1},\dots,G_{\kappa_M}$ are independent and identically distributed given $X$. So, they are exchangeable, and the plus-one $p$-values of Eq.~(6) satisfy $\Pr[p\le\alpha]\le\alpha$~\cite{phipson2016permutation}.
\end{proof}


\paragraph{Consequences.} Attribution also requires $C\ge L$, $\rho=L$, and a unique maximum, so a false attribution has a probability of at most $\alpha+\epsilon$. A positive decision of either kind has a probability of at most $2\alpha+\epsilon$. The bound holds per decision, so a forger that submits $q$ units expects at most $q(\alpha+\epsilon)$ false detections.


\paragraph{Scope.} The bound does not cover reused sessions, whose HMAC inputs repeat inputs in the forger's view. It also does not cover edited copies of authentic logs, whose sessions the forger has seen, or a forger that sees verifier outputs. Condition (A1) needs wrong keys that the forger cannot predict. For reproducibility, our runs derived them from fixed seeds of a non-cryptographic generator. Our forgers never used these keys, so $X$ did not depend on them, but a deployed verifier should draw them with fresh secret randomness.


\paragraph{Relation to Table~\ref{tab:forgery}(a).} The table contains 672 fresh-session decisions over both attackers: 300 single and 180 pooled on ALFWorld, and 120 single and 72 pooled on WebShop-1K. By linearity of expectation, the bound allows at most $672(\alpha+\epsilon)\approx0.67$ expected false detections, even though pooled groups share trajectories. We observed one false detection and no false attribution. A zero-event criterion over this many decisions can fail even when every decision meets $\alpha$, so this event is consistent with the bound. The median single-trajectory $p_{\mathrm{det}}$ of fresh-session forgeries ranged from 0.39 to 0.63 across conditions, close to the 0.5 expected under exchangeability.


\section{Calibration and Cost}
\label{app:extra}


\paragraph{Null calibration.} Clean verification ($M=1999$, $\alpha=10^{-3}$) of unwatermarked evaluation logs gave no false detection or false attribution. This holds for 50 ALFWorld and 30 WebShop-1K single logs, and for 30 and 18 pools from the three partitions. Kolmogorov--Smirnov tests of uniformity gave $p$-values of 0.14 and 0.49 for $p_{\mathrm{id}}$ and 0.89 and 0.44 for $p_{\mathrm{det}}$ on ALFWorld and WebShop-1K. With 50 and 30 logs, these tests are only descriptive.


\paragraph{Cost.} The baseline and watermarked agents compute the same $P_t$. The watermark adds the semantic labels, key derivation, the coding row, and the keyed tilt. On ALFWorld, with one NVIDIA A100 GPU (40\,GB), the mean time per step was 4.62\,s for the baseline and 4.63\,s with the watermark. The loaded model used 17.2\,GB of GPU memory, and the structured log that verification needs took about 1.2\,KB per step.


\section{Key-Input Forgery Experiment}
\label{app:keying}
We report a post-hoc experiment run after the frozen evaluation in Figure~\ref{fig:problem}. It reuses the ALFWorld logs, embedding channel, decoder, wrong-key calibration ($M=999$, $\alpha=10^{-3}$), session identifiers, and seeds of the learned-forgery evaluation in Section~4.2, and it changes only the input of the step key. The AgentMark-style key uses the step index and the three most recent actions, a restricted instance of the context in \cite{huang2026agentmark}. The SeqWM-style key uses the three preceding actions, as in \cite{an2026sequential}. The semantic key is Eq.~(2) without the session identifier, and the AgentAnchor key is Eq.~(2). Forged actions enter the action history that later keys read. For each key, the attacker observes $N\in\{10,50,200\}$ owner-watermarked logs, obtained by re-embedding the 75 unwatermarked base logs under that key, and rewrites the 50 unwatermarked evaluation logs, each bound to an unobserved session identifier. The weak and strong attackers of Section~4 group their observations by the public part of each key, which for AgentAnchor is the semantic window because the target session is fresh. Figure~\ref{fig:problem}(b, c) reports the worst of the two attackers. Genuine watermarked replays of the 50 targets were detected in 48, 47, 45, and 44 cases under the four keys, so the differences do not come from a weaker watermark, and the AgentAnchor arm reproduces the fresh-session counts of Table~\ref{tab:forgery}(a) exactly. At $N=200$, paired exact McNemar tests against AgentAnchor give $p=3\times10^{-5}$, $8\times10^{-3}$, and $2\times10^{-12}$ for the AgentMark-style, SeqWM-style, and semantic keys. The strong attacker changed 84\% of target actions under both the semantic key and AgentAnchor, yet 41 and 1 forgeries passed detection. The semantic key yields few forgeries below $N=200$ because the strong attacker uses a public state only after observing it at least twice. These keys isolate the keying context and do not evaluate AgentMark or SeqWM, which use different samplers, decoders, and thresholds. As in Section~4.2, the observed logs include re-embedded versions of the target tasks, so the absolute rates describe a well-informed attacker.

\end{document}